# New collective states of 2D electrons in high Landau levels

J.P Eisenstein[a], M.P. Lilly[a], K.B. Cooper[a], L.N. Pfeiffer[b] and K.W. West[b]

[a]*California Institute of Technology*
*Pasadena CA 91125 USA*
[b]*Bell Laboratories, Lucent Technologies*
*Murray Hill, NJ 07974 USA*

**ABSTRACT**

A brief summary of the emerging evidence for a new class of collective states of two-dimensional electrons in partially occupied excited Landau levels is presented. Among the most dramatic phenomena described are the large anisotropies of the resistivity observed at very low temperatures near half-filling of the third and higher Landau levels and the non-linear character of the re-entrant integer quantized Hall states in the flanks of the same levels. The degree to which these findings support recent theoretical predictions of charge density wave ground states is discussed and a preliminary comparison to recent transport theories is made.

Corresponding author:
J.P. Eisenstein
Mail Code 114-36
Pasadena, CA 91125
Fax: 626-683-9060
email: jpe@caltech.edu

Twenty years have passed since Klaus von Klitzing's remarkable discovery[1] and yet research on 2D electrons in high magnetic fields continues to flourish. One of the biggest reasons for this is simply that samples keep getting better and better. Today it is possible to obtain, from a number of MBE experts around the world, GaAs heterostructures with low temperature electron mobilities exceeding $10^7 cm^2/Vs$. Considering that the fractional quantized Hall effect, a subtle interaction phenomenon, was discovered[2] using a sample with mobility *one hundred times smaller* than this, it is perhaps not surprising that new discoveries continue to be made. Quite recently a particularly good example of this occurred: Transport measurements on ultra-pure 2D electron systems have uncovered a whole new class of many-electron states which inhabit the excited, as opposed to the ground, Landau level[3-9]. These new states are clearly not the same as the fractional quantized Hall fluids discovered by Tsui, Stormer and Gossard[2] and instead exhibit numerous properties suggestive of charge density waves. This paper reviews some of the main features of these new states.

In large part new collective phenomena in the 2D electron system have been discovered by examining the bumps and wiggles in the magnetic field dependence of the transport coefficients $\rho_{xx}$ and $\rho_{xy}$. The famous $\nu=1/3$ fractional quantized Hall effect was discovered[2] by noting that a deep minimum in $\rho_{xx}$ and a plateau in $\rho_{xy}$ appeared at magnetic field where they weren't expected. The same applies to the case of interest here: anomalies in the resistivity at high Landau level occupancy were noted (as far back as in 1988[9].) that did not seem "right". Figure 1 displays longitudinal resistance data taken at T=150mK from a sample having low temperature mobility of $1.1 \times 10^7 cm^2/Vs$. Above B=2.7T the Fermi level lies in either the N=1 first excited or N=0 ground Landau level (LL). Strong fractional quantum Hall effect (FQHE) states are observed at $\nu$=4/3 and 5/3 in the upper spin branch of the N=0 LL, along with numerous weaker states. In the N=1 LL the even-denominator FQHE states at $\nu$=5/2 and 7/2 are beginning to emerge. The inset shows a blow-up of the resistance in the N=2 LL and reveals a complex set of features. At this high temperature, it is not possible to make much sense of this structure.

Figure 2 shows two sets of longitudinal resistance data from the same sample, only now taken at 25mK. The two traces differ only in the contact configuration used for the measurement. In both cases the current is injected at an ohmic contact at the midpoint of one side of the 5x5mm square sample and is withdrawn at the midpoint contact on the opposite side. The voltage is measured between two corner contacts. As the diagrams suggest, the average current flow directions in the two data sets are perpendicular to one another. For the data shown the sides of the sample are parallel to the natural cleavage directions in GaAs: <110> and <1-10>. For the solid trace, the average current flow is along <110> while for the dashed trace it is along <1-10>. As the figure clearly demonstrates, the results of the two measurements are profoundly different in the magnetic field range between about 1 and 2.8 Tesla.

Numerous similar measurements, involving many high mobility samples taken from several different MBE wafers, have established that the astonishing field-dependent resistance anisotropy apparent in Fig. 2 is not an artifact. The following basic results have been established: A large anisotropy in the resistance of 2D electron systems develops at low temperatures near half-filling of highly excited Landau levels. For current flow in one direction, the resistance exhibits a strong maximum while in the orthogonal direction a minimum is observed. The effect is largest at $\nu$=9/2 and 11/2 in the N=2 third Landau level but persists, with decreasing strength, into several higher LLs. The attenuation of the anisotropy with increasing filling factor is generally monotonic, but contains a noticeable oscillatory dependence upon the spin sublevel. In all samples that we have studied the "hard" transport direction is roughly along <1-10> and the "easy" direction is along <110>, provided that no in-plane magnetic field component[5] is added to the existing perpendicular field. These principal axes appear to be unaffected by changes in the definition of the sample boundaries, at least in mm-size samples. No significant anisotropy is seen in the N=1 LL at $\nu$=5/2 or 7/2 (again in the absence of an in-plane magnetic field) or in the ground N=0 LL. The high Landau level anisotropy is not observed above about 150mK but, as Fig. 3 demonstrates, turns on rapidly as the temperature is reduced below 100mK. When fully developed the effect is largest at half-filling but remains substantial in a window of about $\Delta\nu\approx0.2$ about this point. The width of the main peak in the hard resistance near half filling remains roughly constant as the temperature is reduced, in contrast to the behavior expected in the conventional model of a disorder-driven transitions between integer quantized Hall states. In the flanks of the N≥2 Landau levels the resistivity becomes essentially isotropic but exhibits a remarkable *re-entrance* of the integer QHE.

At $\nu$=9/2 resistance anisotropy ratios of 100 are common in square samples with mobilities of $10^7$cm$^2$/Vs or higher. It has been pointed out, however, that current spreading effects in square samples exaggerate the intrinsic microscopic resistivity anisotropy between $\rho_{xx}$ and $\rho_{yy}$[10]. Results of experiments on Hall bar samples have verified this, but do not alter the basic conclusion that transport near half-filling of high

Landau levels is highly anisotropic. Figure 4 shows data obtained from two Hall bars cut from the same parent MBE wafer. In one case (solid curve) the long axis of the bar is along <110> while in the other (dashed curve) it is along <1-10>. The data in the figure, taken at T=20mK, show a roughly 10-fold anisotropy of the resistance near $\nu=9/2$ and a comparable effect at $\nu=11/2$. This observation, as well as a careful comparison of the temperature development of the anisotropy in Hall bars and square samples offers strong evidence that both geometries reveal the same qualitative effect.

Two other aspects of the data in Fig. 4 are worth noting. First, the resistance peak between $\nu=4$ and $\nu=5$ is asymmetric and not centered exactly at $\nu=9/2$. Second, the resistance peak at $\nu=11/2$ is higher than that at $\nu=9/2$. These features are unusual but serve as reminders that there are variations in the detailed transport characteristics from one sample to the next.

The rapid development, at very low temperatures, of large anisotropies in the longitudinal resistance of high mobility 2D electron systems might suggest that something dramatic is happening to the ground state of the system. On the other hand, a more mundane explanation, based upon a weak anisotropic scattering mechanism that becomes important at very low temperature might also be possible. In our opinion, observed filling factor dependence of the anisotropy strongly favors a ground state effect. As Fig. 2 shows, the 100-fold anisotropy at $\nu=9/2$ disappears almost entirely when the magnetic field is raised by 0.7T to reach $\nu=7/2$. Studies of several samples, with 2D densities varying by about a factor of 2, have consistently shown that the onset of anisotropy is keyed to the filling factor and not the magnetic field itself. In addition, the anisotropy gradually disappears in the semiclassical regime at very low magnetic field.

These facts are difficult to reconcile within an anisotropic scattering model. On the other hand, the detailed structure of the Landau level wavefunction is different in each level (e.g. the number of nodes in the wavefunction equals the level index N) and such differences can qualitatively affect the electronic ground state. For example, fractional quantum Hall states exist at half-filling (at $\nu=5/2$ and $7/2$) of the N=1 LL while in the N=0 lowest LL the ground states at $\nu=1/2$ and $3/2$ are compressible composite fermion metals. Ultimately, this difference arises from the different structure of the Landau level wavefunction.

In 1996, two years prior to the experiments of Lilly, et al.[3] and Du, et al.[4], Koulakov, Fogler and Shklovski[11] and Moessner and Chalker[12] predicted, on the basis of a Hartree-Fock (HF) analysis, that the ground state of a 2D electron gas at high Landau level occupancy would be a charge density wave (CDW). This conclusion contrasts with the situation in the lowest Landau level, where various strongly correlated uniform density quantum liquids (i.e. the Laughlin states and related higher order fractions plus the composite fermion metals[13]) are lower in energy than HF CDWs[14]. The additional nodes in the single particle wavefunctions in high LLs alter the balance between the exchange energy, which prefers a higher electron density, and the direct Coulomb repulsion that favors a uniform state. In essence, the nodes soften the repulsion at short distances sufficiently that the exchange effect wins. The result is phase separation on a length scale comparable to the classical cyclotron radius $R_c=\ell\nu^{1/2}$ (where $\ell=(\hbar/eB)^{1/2}$ is the magnetic length and $\nu$ is the filling factor). At half filling of a high Landau level the CDW is expected to be a unidirectional stripe phase with wavelength $\lambda\approx 3R_c$. At $\nu=9/2$, for example, stripes of filling factor $\nu=4$ alternate with stripes of $\nu=5$.

Moving away from half filling one set of stripes widens while the other narrows. Eventually a transition to a triangular CDW composed of multi-electron "bubbles" is predicted to occur. Deep in the flanks of the level this bubble phase becomes equivalent to a conventional Wigner crystal.

This CDW scenario offers a qualitative explanation for many of the experimental results discussed above. It is certainly plausible that the unidirectional CDW expected near half filling would exhibit anisotropic transport if the stripe orientation was pinned coherently across the sample. Anisotropy would still be expected even if the stripe pattern was broken into domains, provided that on average they prefer a particular orientation. At the same time, the observation that the resistivity becomes isotropic in the flanks of the Landau levels is consistent with the expected transition to a triangular CDW phases. Finally, and perhaps most remarkable, is the agreement between theory and experiment that the filling factor boundary separating low Landau level isotropic transport from high Landau level anisotropy ) is at $\nu=4$, i.e. at the transition between the N=1 and N=2 LLs.

Following the initial experimental results of Lilly, et al.[3] and Du, et al.[4] a flurry of theoretical activity ensued. Numerical exact diagonalization calculations by Haldane, Rezayi, and Yang[15] lent significant weight to the early Hartee-Fock results. These new calculations clearly revealed strong peaks in the susceptibility and structure factor of 2D electrons at half filling of the $N\geq2$ LLs. The wavevector of these peaks is consistent with the HF result of $\sim3R_c$ for the stripe wavelength. Fradkin and Kivelson[16] examined the role of quantum fluctuations on the CDW ordering and argued that if strong enough the CDW would be melted and behave as a nematic liquid crystal. Fertig[17] argued that simple 1D stripes are in fact unstable to density modulations along their length but the resulting anisotropic Wigner crystal is likely melted. Stanescu, et al[18] and Jungwirth, et al.[19] examined the effect of in-plane magnetic fields on the stripe phase in an effort to understand the newer experimental results of Lilly, et al.[5] and Pan, et al.[6] which show that the principal axes of the anisotropy can be drastically altered by tilting the sample slightly in the magnetic field. Recently, Fogler and Vinokur[20] and Coté and Fertig[21] have considered the hydrodynamics and collective modes of the stripe phases.

We emphasize that so far only macroscopic transport measurements have been performed. Such measurements do not prove the existence of a microscopic modulation of the charge density in the system. Until more incisive probes are applied to the high Landau level problem, the case for stripes is essentially circumstantial. On the other hand, transport theories of striped quantum Hall systems have begun to emerge and they offer the prospect of quantitative confrontation between theory and experiment. MacDonald and Fisher[22] (MF) have examined a model of parallel stripes in which transport is governed by the hopping rates, due to impurity scattering, of electrons between the edge states which run along the stripe boundaries. Using this model MF were able to deduce a number of interesting results. For example, the Luttinger liquid aspects of the edge states leads naturally to a non-linearity of the resistivity. Although we have not discussed it here, Lilly, et al.[3] reported that a significant non-linearity in the resistivity develops under the same conditions as the anisotropy. MF also showed that within their model certain *universal* relationships between the transport parameters hold. For example, they find that the product $\rho_{xx}\rho_{yy}$ of the microscopic resistivities at half-filling is totally independent of the hopping rates:

$$\rho_{xx}\rho_{yy} = (h/e^2)^2 \frac{1}{\left[(1+[\nu])^2 + [\nu]^2\right]^2} \tag{1}$$

where [ν] = int(ν). Not surprisingly, no similar result exists for the interesting anisotropy ratio $\rho_{xx}/\rho_{yy}$; this depends on the scattering rates. In order to compare Eq. 1 to experimental data, the geometric effects connecting the measured resistances to the desired resistivities must be accounted for. Simon[23] has analyzed this problem using a classical model of the current distribution within an anisotropic Hall conductor with contacts arranged after the manner of Lilly, et al.[3]. This analysis allows for a conversion between measured resistances, such as those in Fig. 2, and the underlying resistivities. We emphasize that such a conversion assumes at the outset that the many complicating effects characteristic of transport in quantum Hall devices (e.g. non-local effects, disequilibrium between edge and bulk channels, etc.) can be ignored. In spite of this, Fig. 5 reveals that the comparison of Eq. 1 and the resistivity values extracted from the experimentally measured resistances at ν=9/2, 11/2, etc. in Fig. 2 is remarkably good.

The model used by MF consists of uniform parallel stripes. A related analysis by von Oppen, Halperin and Stern (vOHS)[23] has shown that Eq. 1 applies to much more general situations in which there is substantial disorder in the stripe pattern. Indeed, in the vOHS picture Eq. 1 continues to hold even when the stripe pattern is so disordered that the macroscopic resistance is rendered *isotropic*. Thus, the good agreement between theory and experiment shown in Fig. 5 does not, by itself, suggest a uniform array of well-ordered stripes having few defects. In addition, vOHS have generalized Eq. 1 to filling factors away from half-filling. They find a "semicircle law" connecting the dissipative and Hall conductivities:

$$\sigma_{xx}\sigma_{yy} + (\sigma_{xy} - \sigma_0)^2 = (e^2 / 2h)^2 \qquad (2)$$

In this formula $\sigma_0$ is the classical Hall conductivity expected at the half-filling point (e.g. for $4<\nu<5$, $\sigma_0=9e^2/2h$, etc.). Figure 6 compares Eq. 2 with experimental results from the filling factor range $4<\nu<6$, covering the anisotropic states at $\nu=9/2$ and 11/2. As in Fig. 5, the measured resistances were first converted into resistivities using the Simon analysis[10]. The conductivities were then obtained by inverting the resistivity tensor. The comparison between theory and experiment is again surprisingly good. We stress, however, that these comparisons with theory, while interesting and encouraging, are preliminary. They rely heavily on the Simon model of the current distribution and are subject to sample-to-sample variations. Furthermore, it is worth remembering that in the standard model[24] of localization-driven integer QHE transitions (in stripe-free isotropic systems), both Eq. 1 and 2 are valid, albeit in a singular way, at zero temperature. That they seem to work in a strongly anisotropic system in which the width of the resistivity features around half-filling appears to remain finite as $T\rightarrow0$ is remarkable.

The anisotropic transport features near half-filling are not the only interesting effects seen in high Landau levels. Figure 7 shows both longitudinal and Hall resistances in the range $4<\nu<5$. Near $\nu=9/2$ the large longitudinal resistance anisotropy is accompanied by a smooth and nearly featureless Hall resistance. Away from half filling, roughly at $\nu\approx4.25$ and 4.75, something different is apparent: The longitudinal resistances are both tending to zero, suggesting that fractional quantized Hall states may be present. Indeed, the Hall resistance is quantized, but at the nearby *integer* value. The blow-ups in the figure display the effect around $\nu\approx4.25$ clearly. The broad $\nu=4$ integer Hall plateau is interrupted by a narrow non-quantized region before returning to quantization near $\nu \approx 4.25$. The deviation from quantization coincides with the small peak in the longitudinal resistance separating the deep $\nu=4$ zero from the weaker one at $\nu \approx 4.25$. This re-entrance of the integer QHE is seen most clearly in the N=2 LL, but hints have been detected in higher LLs. Like the anisotropy, it is not found in the N=1 or N=0 LLs.

The existence of these re-entrant integer QHE states (RIQHE) reflects underlying insulating configurations for the quasiparticles. That they are separated, at finite temperature, from the main integer QHE states suggests a different kind of localization is operative. In the conventional view of the integer QHE the localization is a single particle effect. Here it would seem that a more collective phenomenon should be considered.

To examine this possibility, we have studied the non-linear behavior of the longitudinal resistance in the re-entrant integer QHE states. Figure 8 shows typical dc I-V characteristics from two samples taken at T=25mK in the center of the re-entrant state near $\nu \approx 4.25$. Abrupt and hysteretic transitions from insulating to conducting states are seen in both samples. Similar transitions are found at the $\nu \approx 4.75$ RIQHE state. Figure 9 shows a family of such I-V characteristics taken at several magnetic fields from the low field side of the $\nu \approx 4.25$ state upwards to $\nu=4$. The conventional longitudinal resistance is plotted alongside for reference. Note that the discontinuous behavior is found only in the re-entrant QHE region. As reported by Cooper, et al.[7], the discontinuous behavior is restricted to very low temperatures. At the center of the $\nu \approx 4.25$ RIQHE the I-V jumps persist up to about 60mK. This critical temperature falls on moving away from the center of the RIQHE.

This non-linear behavior is reminiscent of depinning and sliding conduction in conventional CDW compounds like $NbSe_3$[25]. In the present case, the current flowing through the sample (via the filled LLs beneath the Fermi level) produces a Hall electric field that exerts a force, transverse to the current, on the localized quasiparticles in the uppermost LL. For small currents this force is unable to delocalize these quasiparticles and so $R_{xx}$ and $R_{yy}$ remain zero. Eventually, however, the force becomes large enough that delocalization occurs and non-zero longitudinal voltages appear. If the electrons were initially individually localized and acted independently, a smooth transition to conduction might be expected. On the other hand, the observed discontinuous nature of the transition suggests a more collective effect.

The early Hartree-Fock calculations of Koulakov, Fogler, and Shklovskii[11] and Moessner and Chalker[12] conclude that triangular "bubble phase" CDWs exist in the flanks of high Landau levels. Depending on the precise filling factor and the Landau level index, the number of electrons per bubble can be 1,2,3 or more. In the N=2 LL only one- and two-electron bubbles are expected[11]. Recent exact diagonalization calculations support this conclusion[15]. Such CDW's are likely pinned by disorder and thereby leave $R_{xx}=R_{yy}=0$. The discontinuous transitions to conduction that we observe might therefore represent the depinning and sliding conduction of these CDW's. However, before other mechanisms (e.g. run-away heating[26] and other QHE breakdown phenomena[27]) can be ruled out, more definitive evidence (e.g. observation of narrow band noise) is needed.

In conclusion, we have reviewed the recent evidence that truly new many-electron states exist in highly excited Landau levels at very low temperatures. Giant anisotropies in the longitudinal resistance near half filling and curious re-entrant integer QHE states in the flanks of these levels are among the most striking new findings. Other intriguing aspects, notably the strong sensitivity of the anisotropic states to added in-plane magnetic fields are also seen but space did not permit their inclusion here. While there is considerable overlap between the experimental results and the growing theoretical literature on charge density wave states, the evidence remains largely circumstantial. There is much room for further work.

We thank B. Halperin, A.H. MacDonald, E. Rezayi, A. Stern, and K. Yang for enlightening discussions of their most recent work prior to its publication. We are indebted to the Department of Energy and the National Science Foundation for their support.

## FIGURE CAPTIONS

**Fig. 1** Overview of the longitudinal resistance of a high mobility 2D electron system at T=150mK. Inset expands data from N=2 Landau level.

**Fig. 2** Anisotropy of longitudinal resistance at T=25mK. Dashed curve: average current flow along <1-10>; solid curve: average current flow along <110>.

**Fig. 3** Longitudinal resistances vs. temperature at $\nu=9/2$.

**Fig. 4** Longitudinal resistances vs. filling factor measured using two perpendicular Hall bars. Dashed curve: Hall bar axis along <1-10>; solid curve: Hall bar axis along <110>.

**Fig. 5** Resistivity product $\rho_{xx}\rho_{yy}$ at $\nu=9/2$, 11/2, etc. Open circles: MacDonald-Fisher theory[22]; solid circles: Experimental data (T=25mK), after converting resistances to resitivities using Simon analyis[10].

**Fig. 6** Comparison to "semicircle law" of von Oppen, Halperin, and Stern[23]. Dashed curves: theory. Solid curve: Experimental data (T=25mK), after converting resistances to resitivities using Simon analyis[10] and then inverting tensor to obtain conductivities.

**Fig. 7** (a) Longitudinal (solid line: $R_{xx}$, dotted line: $R_{yy}$) and Hall ($R_{xy}$) resistance between $4<\nu<5$ at T=50mK. Insets (b) and (c) magnify region showing re-entrant IQHE near $\nu\approx4.25$.

**Fig. 8** Typical I-V characteristics observed in RIQHE near $\nu\approx4.25$ at T=25mK.

**Fig. 9** Left panel: Family of I-V curves at several closely spaced magnetic fields. Right panel: Longitudinal resistance $R_{yy}$ over same field range. Discontinuous I-V curves are only seen inside RIQHE.

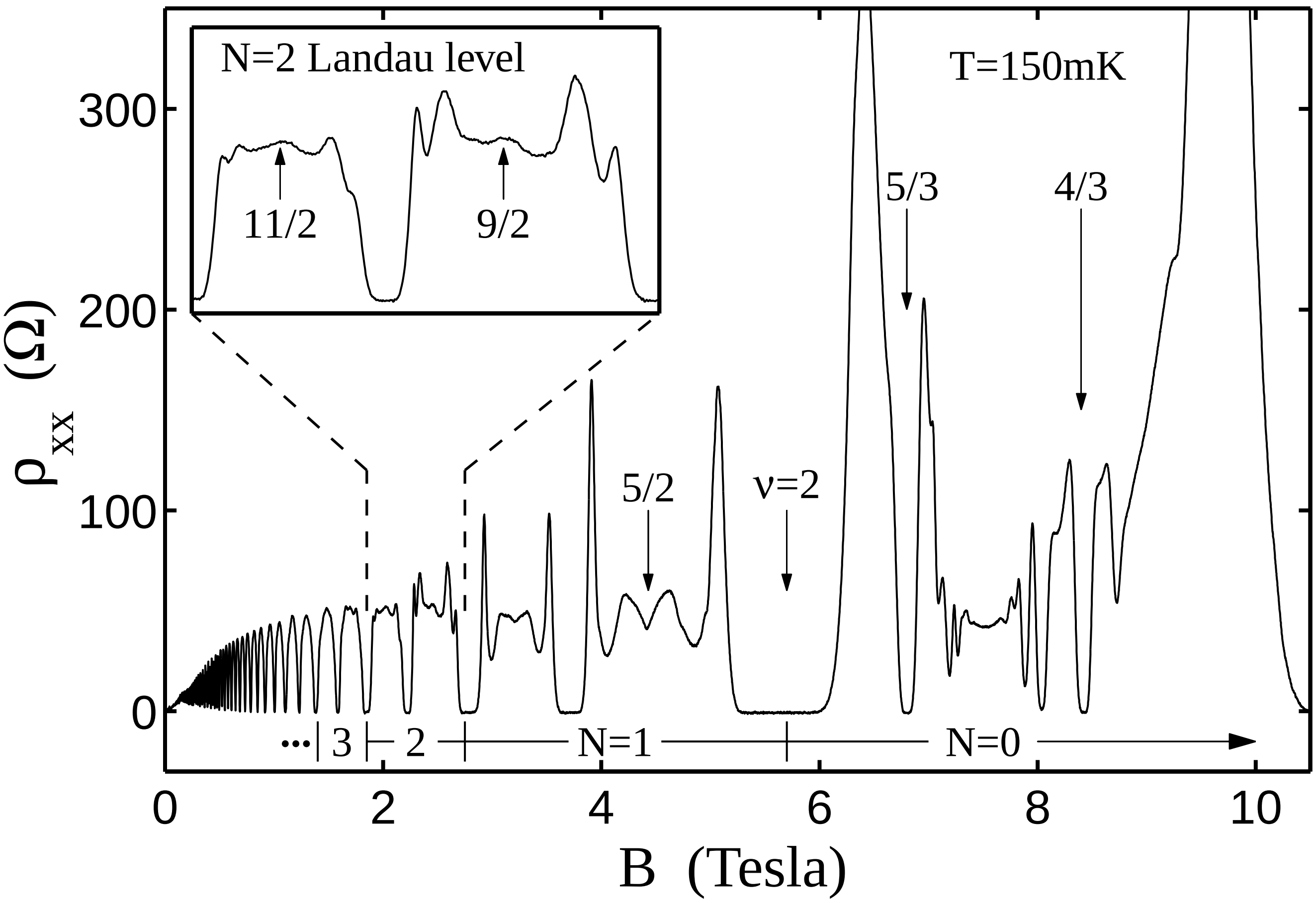

Figure 1

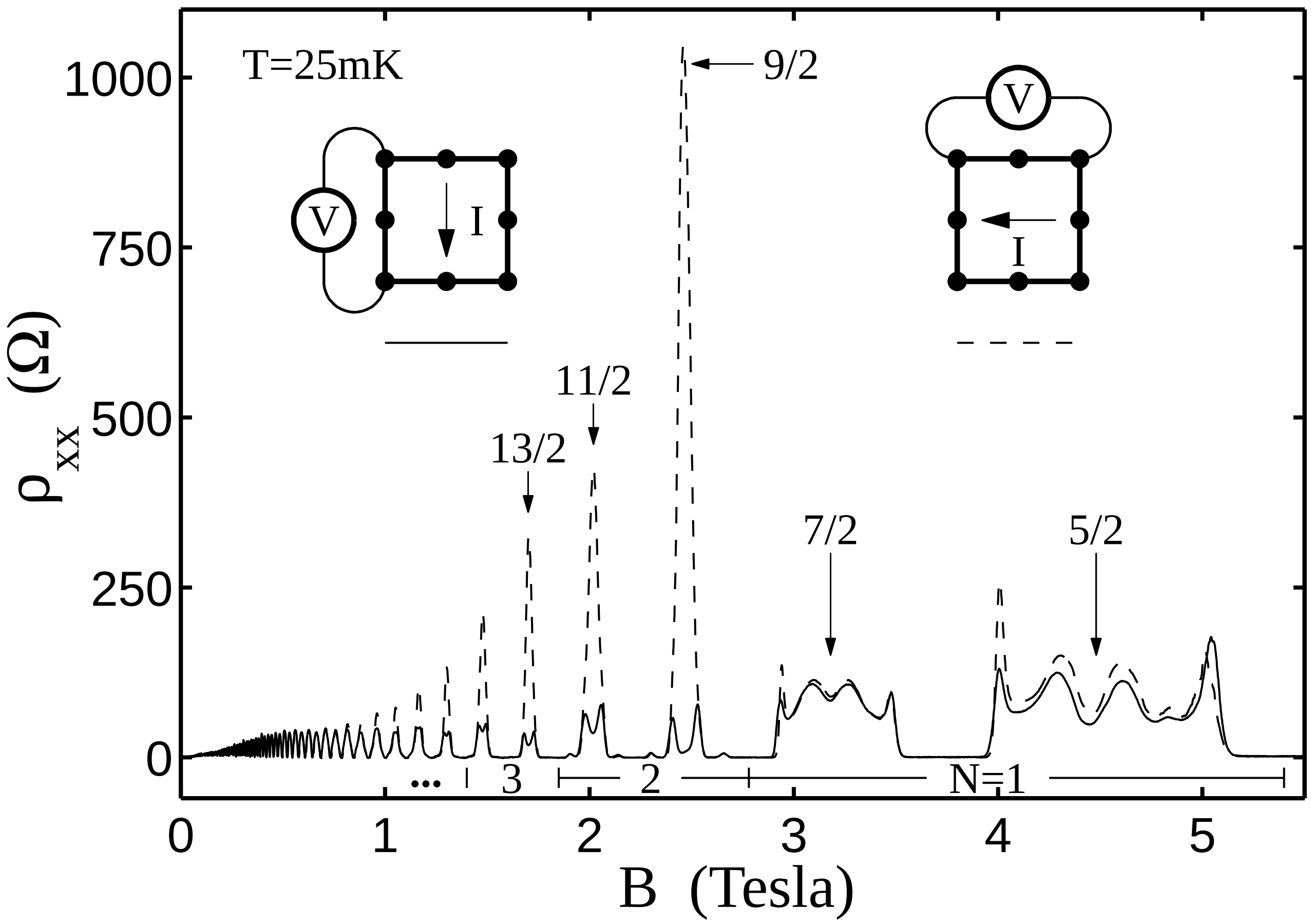

Figure 2

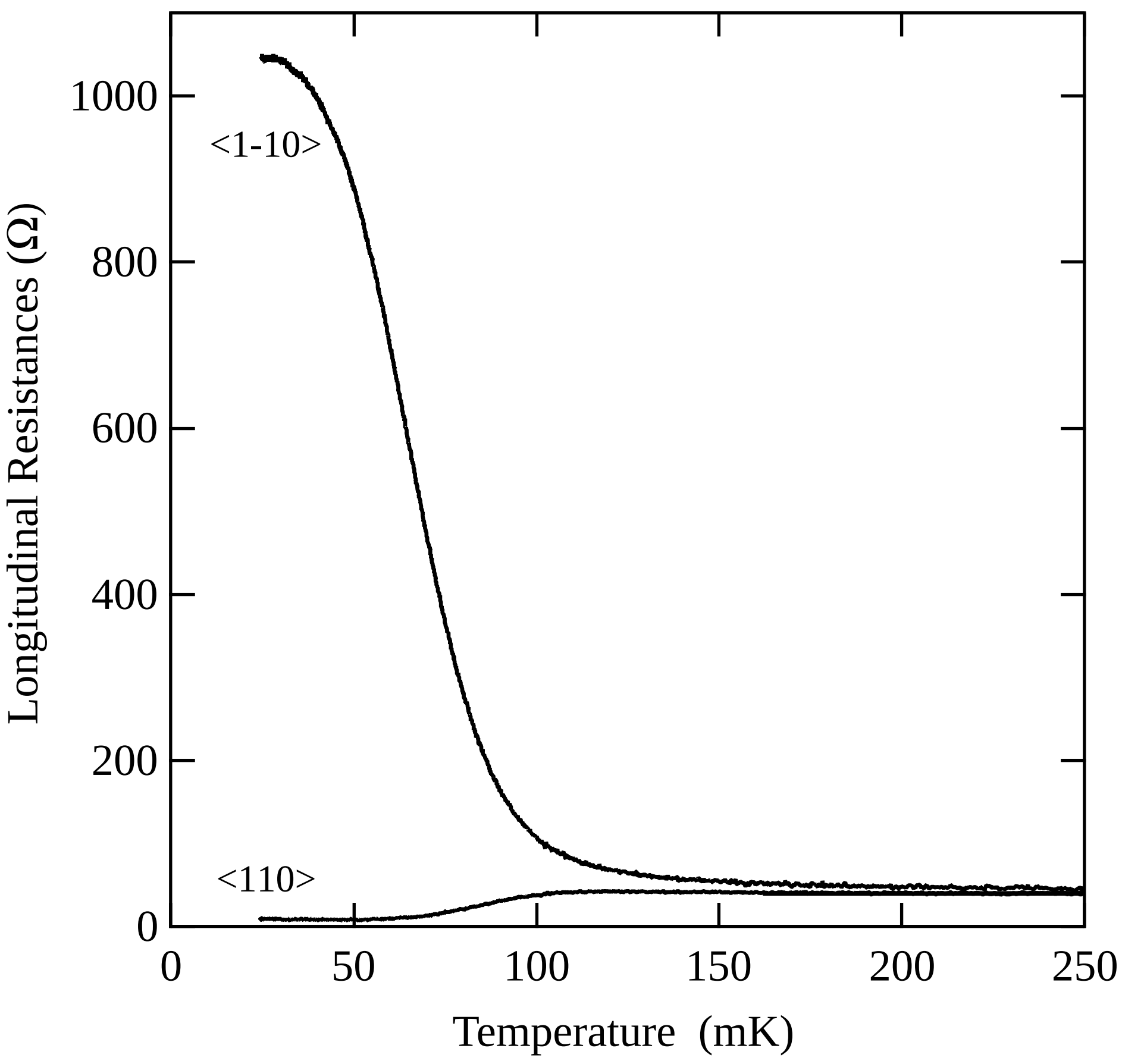

Figure 3

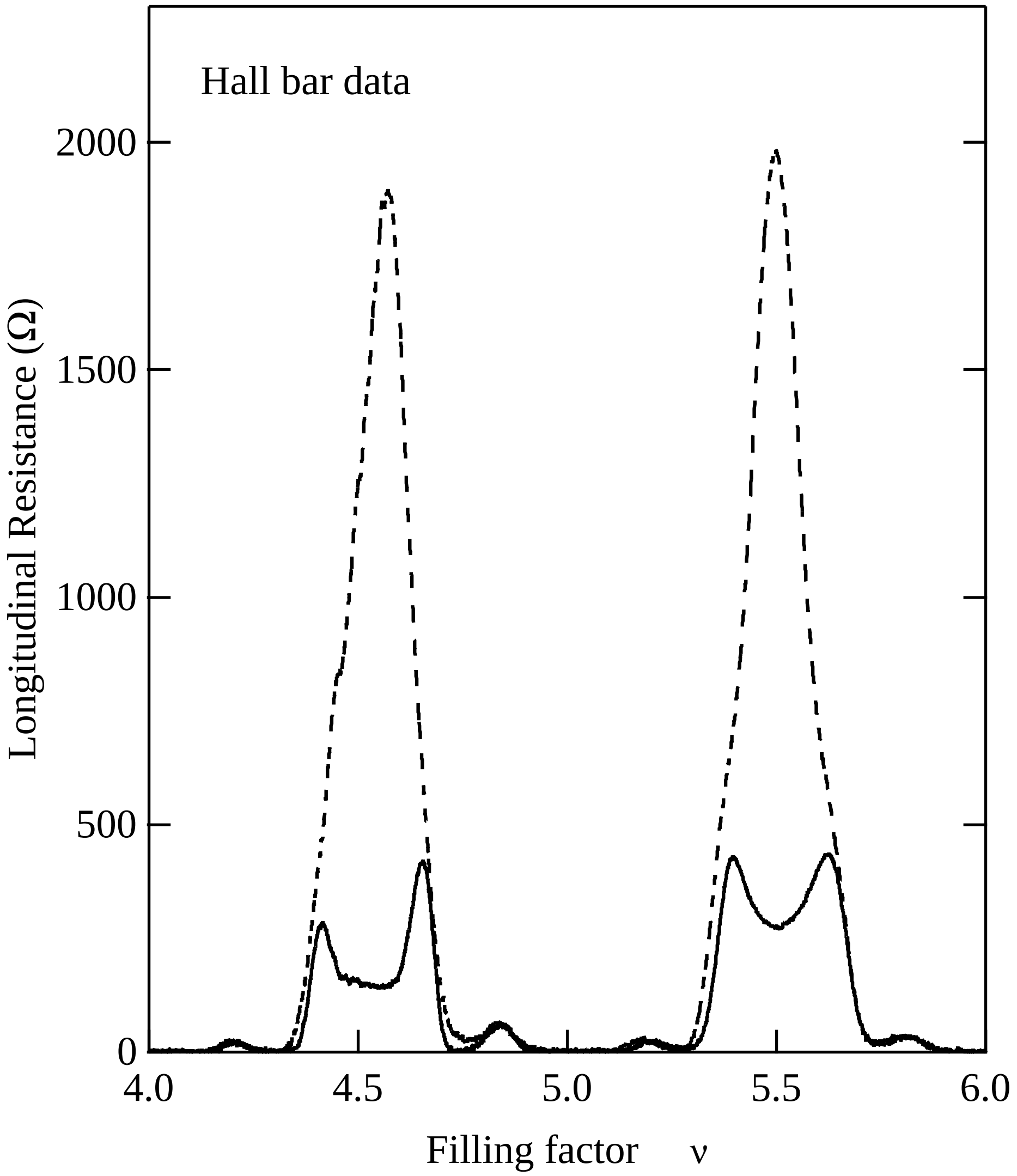

Figure 4

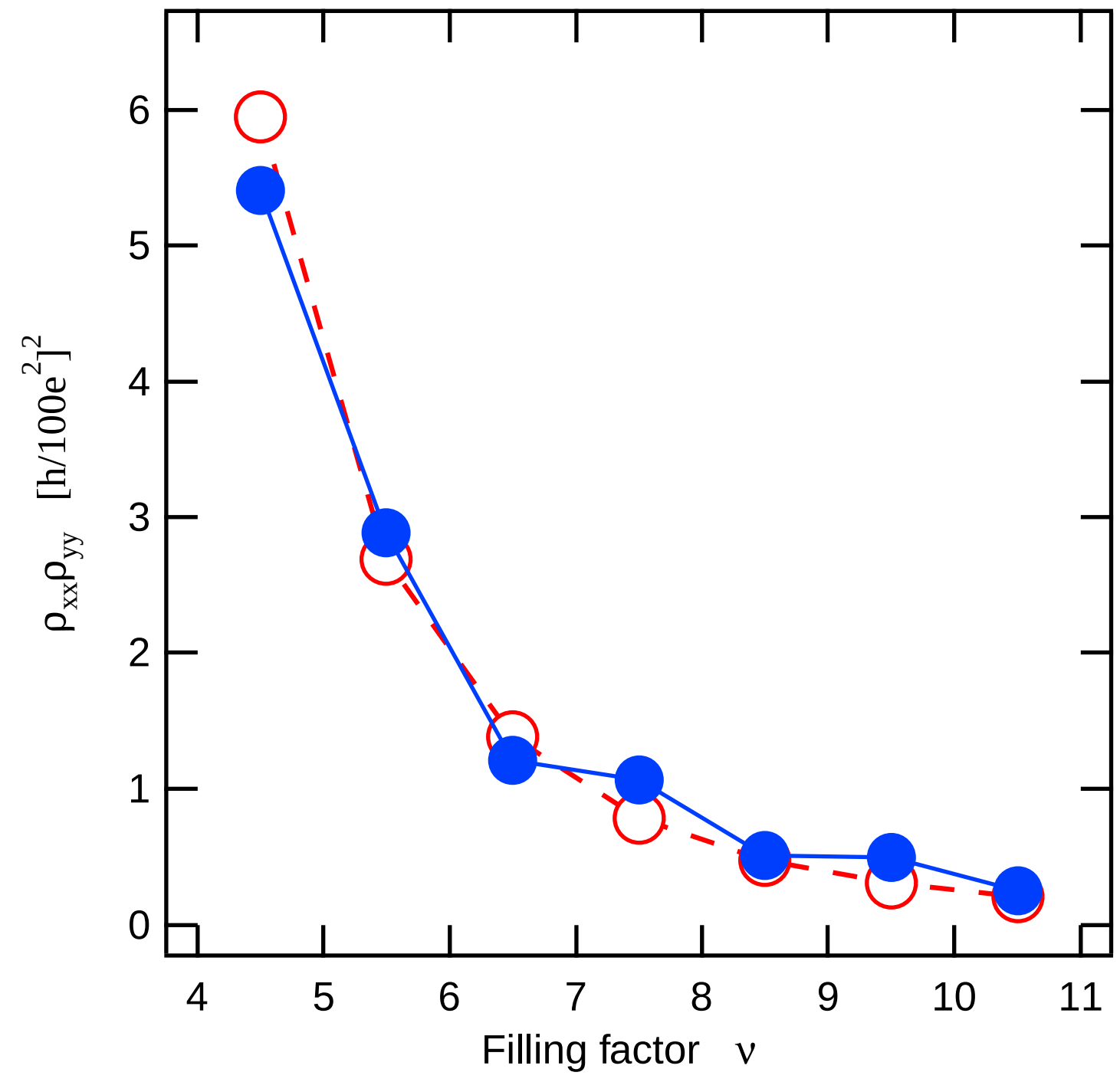

Figure 5

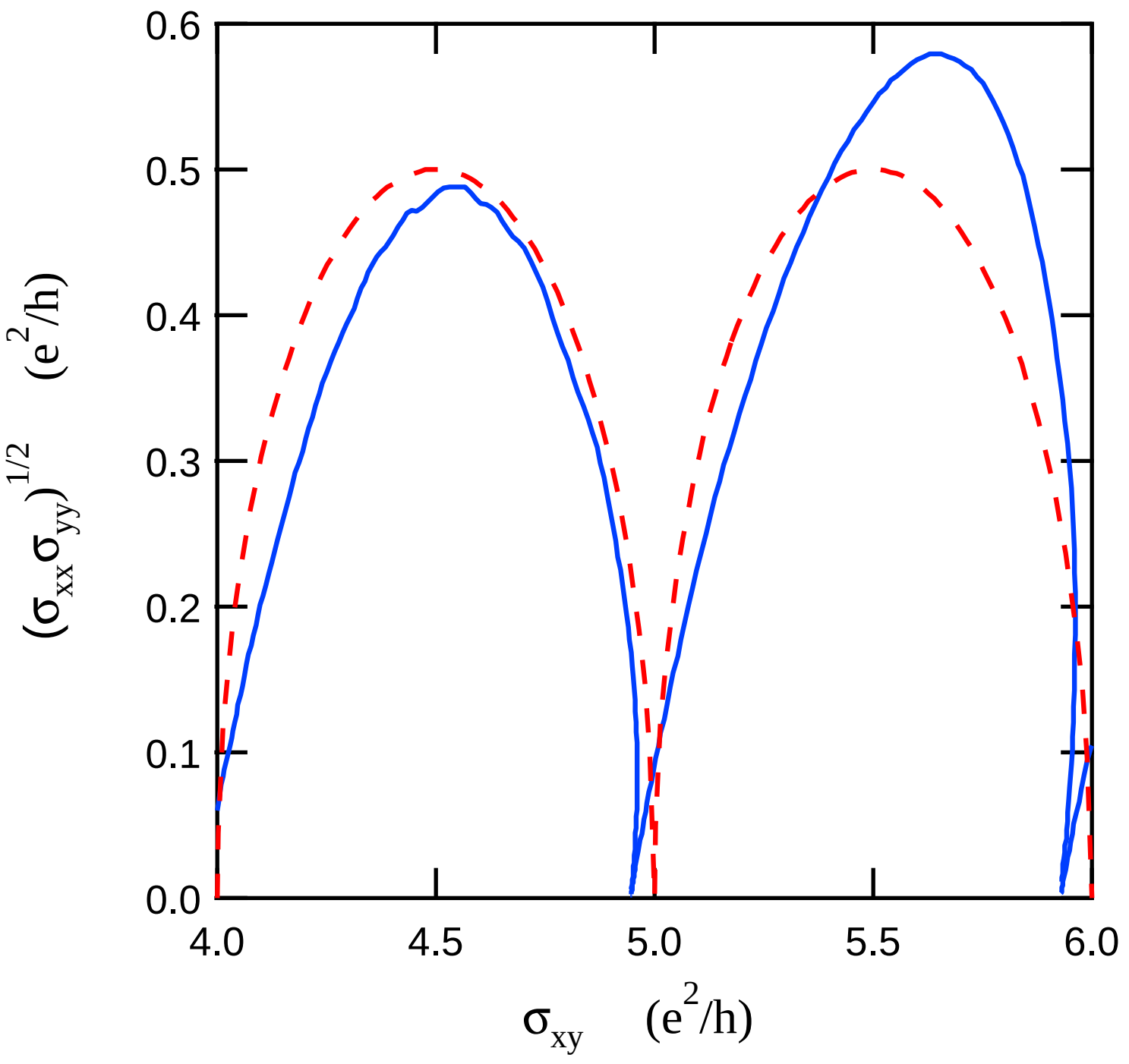

Figure 6

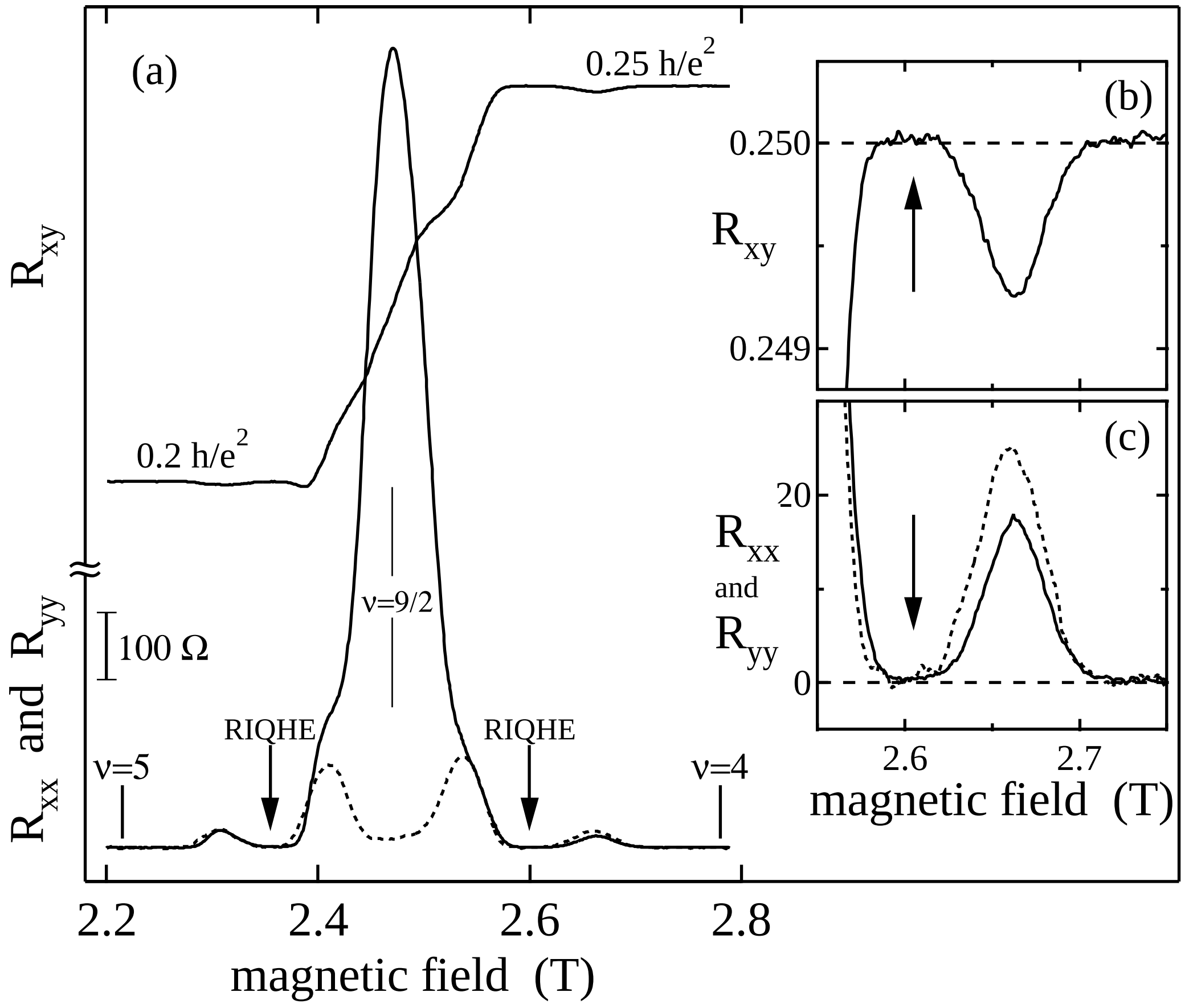

Figure 7

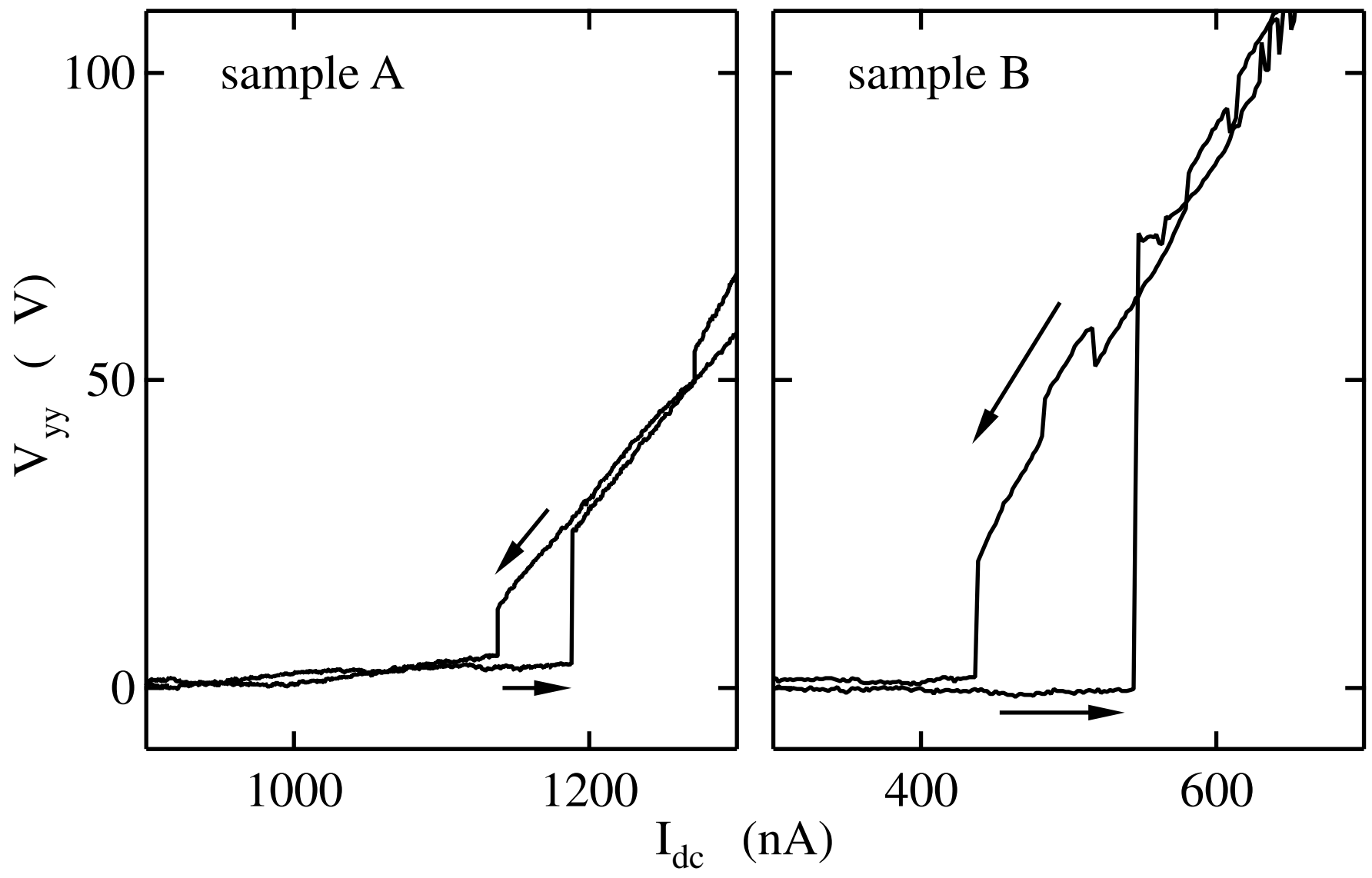

Figure 8

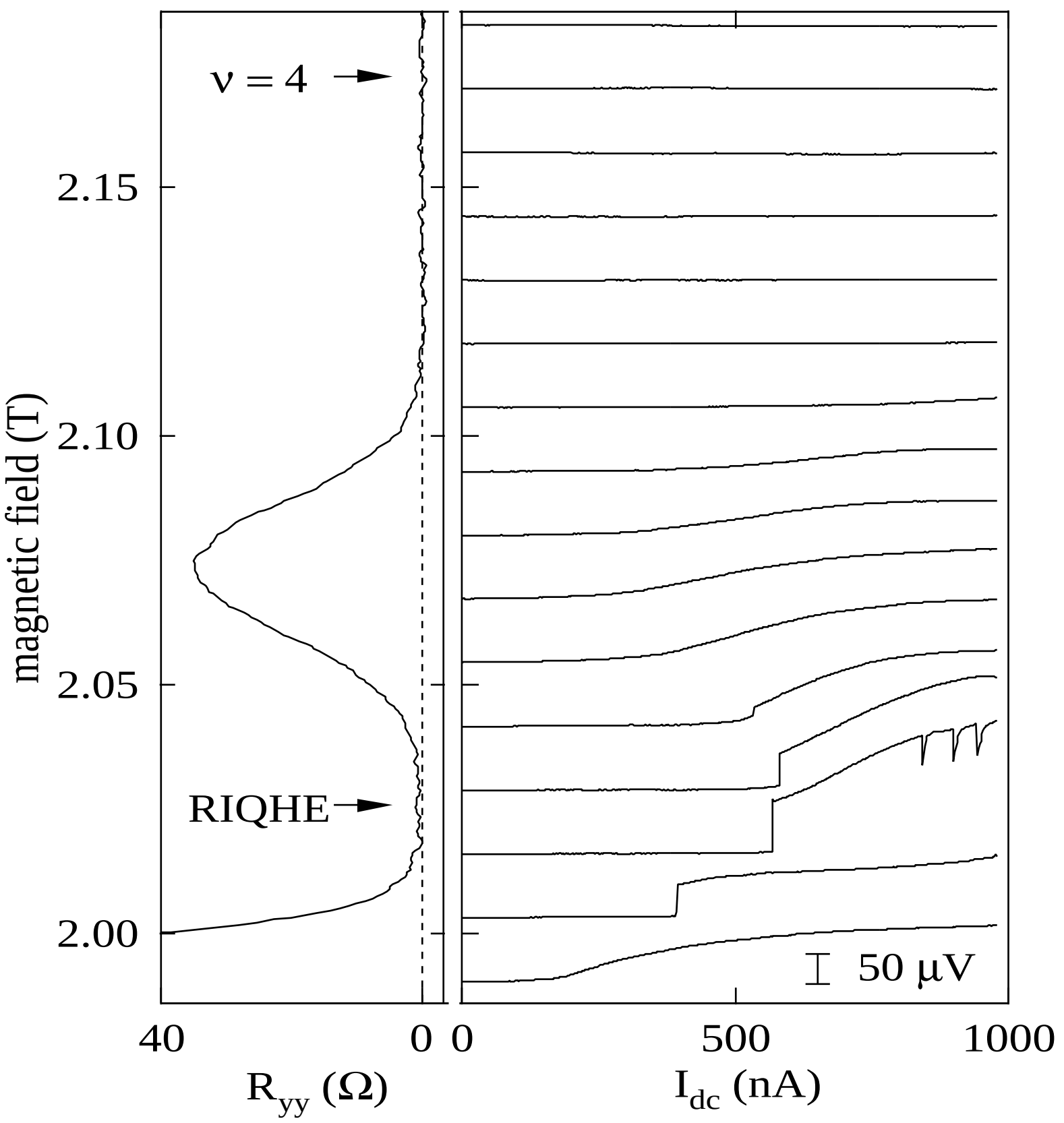

Figure 9